\documentclass[aps,prd,twocolumn,showpacs,nofootinbib,showkeys]{revtex4-1}

\usepackage{amssymb} \usepackage{amsmath} \usepackage{graphicx}
\usepackage{epsfig,latexsym}

\RequirePackage{xspace} \allowdisplaybreaks

\begin{document}

\def\bef{\begin{figure}}
\def\eef{\end{figure}}

\newcommand{\nl}{\nonumber\\}

\newcommand{\ans}{ansatz }
\newcommand{\be}[1]{\begin{equation}\label{#1}}
\newcommand{\beq}{\begin{equation}}
\newcommand{\ee}{\end{equation}}
\newcommand{\beqn}[1]{\begin{eqnarray}\label{#1}}
\newcommand{\eeqn}{\end{eqnarray}}
\newcommand{\bd}{\begin{displaymath}}
\newcommand{\ed}{\end{displaymath}}
\newcommand{\mat}[4]{\left(\begin{array}{cc}{#1}&{#2}\\{#3}&{#4}
\end{array}\right)}
\newcommand{\matr}[9]{\left(\begin{array}{ccc}{#1}&{#2}&{#3}\\
{#4}&{#5}&{#6}\\{#7}&{#8}&{#9}\end{array}\right)}
\newcommand{\matrr}[6]{\left(\begin{array}{cc}{#1}&{#2}\\
{#3}&{#4}\\{#5}&{#6}\end{array}\right)}
\newcommand{\cvb}[3]{#1^{#2}_{#3}}
\def\lsim{\raise0.3ex\hbox{$\;<$\kern-0.75em\raise-1.1ex
e\hbox{$\sim\;$}}}
\def\gsim{\raise0.3ex\hbox{$\;>$\kern-0.75em\raise-1.1ex
\hbox{$\sim\;$}}}
\def\abs#1{\left| #1\right|}
\def\simlt{\mathrel{\lower2.5pt\vbox{\lineskip=0pt\baselineskip=0pt
           \hbox{$<$}\hbox{$\sim$}}}}
\def\simgt{\mathrel{\lower2.5pt\vbox{\lineskip=0pt\baselineskip=0pt
           \hbox{$>$}\hbox{$\sim$}}}}
\def\unity{{\hbox{1\kern-.8mm l}}}
\newcommand{\eps}{\varepsilon}
\def\ep{\epsilon}
\def\ga{\gamma}
\def\Ga{\Gamma}
\def\om{\omega}
\def\omp{{\omega^\prime}}
\def\Om{\Omega}
\def\la{\lambda}
\def\La{\Lambda}
\def\al{\alpha}
\newcommand{\ov}{\overline}
\renewcommand{\to}{\rightarrow}
\renewcommand{\vec}[1]{\mathbf{#1}}
\newcommand{\vect}[1]{\mbox{\boldmath$#1$}}
\def\tm{{\widetilde{m}}}
\def\mcirc{{\stackrel{o}{m}}}
\newcommand{\Dm}{\Delta m}
\newcommand{\dm}{\varepsilon}
\newcommand{\tanb}{\tan\beta}
\newcommand{\nbar}{\tilde{n}}
\newcommand\PM[1]{\begin{pmatrix}#1\end{pmatrix}}
\newcommand{\up}{\uparrow}
\newcommand{\down}{\downarrow}
\def\omE{\omega_{\rm Ter}}
%
%%%%%%%%%%     mauri    %%%%%%%%%%%%%%%%%%%%%%%%%%%%%%%%%

\newcommand{\Dsusy}{{susy \hspace{-9.4pt} \slash}\;}
\newcommand{\DCP}{{CP \hspace{-7.4pt} \slash}\;}
\newcommand{\mc}{\mathcal}
\newcommand{\gr}{\mathbf}
\renewcommand{\to}{\rightarrow}
\newcommand{\gtc}{\mathfrak}
\newcommand{\wh}{\widehat}
\newcommand{\br}{\langle}
\newcommand{\kt}{\rangle}

%%%%%%%%%%%%%%%%%%%%%%%%%%%%%%%%%%%%%%%%%%%%%%%%%%%%%%%%%%

% barbara Ricci  %definizione di minore e maggiore simile
\def\lsim{\mathrel{\mathop  {\hbox{\lower0.5ex\hbox{$\sim$}
\kern-0.8em\lower-0.7ex\hbox{$<$}}}}}
\def\gsim{\mathrel{\mathop  {\hbox{\lower0.5ex\hbox{$\sim$}
\kern-0.8em\lower-0.7ex\hbox{$>$}}}}}
%%%%%%%%%%%%%%%%%%%%%%%%%%%%%%%%%%

\def\nn{\\  \nonumber}
\def\de{\partial}
\def\brf{{\mathbf f}}
\def\bbf{\bar{\bf f}}
\def\bF{{\bf F}}
\def\bbF{\bar{\bf F}}
\def\bA{{\mathbf A}}
\def\bB{{\mathbf B}}
\def\bG{{\mathbf G}}
\def\bI{{\mathbf I}}
\def\bM{{\mathbf M}}
\def\bY{{\mathbf Y}}
\def\bX{{\mathbf X}}
\def\bS{{\mathbf S}}
\def\bb{{\mathbf b}}
\def\bh{{\mathbf h}}
\def\bg{{\mathbf g}}
\def\bla{{\mathbf \la}}
\def\bmu{\mathbf m }
\def\by{{\mathbf y}}
\def\bmu{\mbox{\boldmath $\mu$} }
\def\bsig{\mbox{\boldmath $\sigma$} }
\def\bunity{{\mathbf 1}}
\def\cA{{\cal A}}
\def\cB{{\cal B}}
\def\cC{{\cal C}}
\def\cD{{\cal D}}
\def\cF{{\cal F}}
\def\cG{{\cal G}}
\def\cH{{\cal H}}
\def\cI{{\cal I}}
\def\cL{{\cal L}}
\def\cN{{\cal N}}
\def\cM{{\cal M}}
\def\cO{{\cal O}}
\def\cR{{\cal R}}
\def\cS{{\cal S}}
\def\cT{{\cal T}}
\def\eV{{\rm eV}}
%
%%%%%%%%%%%%%%%%%%%%%%%%%%%%%%%%%%%%%

\title{Limiting Majoron self-interactions from Gravitational waves experiments}

\author{Andrea Addazi$^1$}\email{3209728351@qq.com}
\author{Antonino Marcian\`o$^1$}\email{marciano@fudan.edu.cn}
\affiliation{$^1$ Department of Physics \& Center for Field Theory and Particle Physics, Fudan University, 200433 Shanghai, China}

\begin{abstract}

We show how Majoron models may be tested/limited in gravitational waves experiments. In particular, the Majoron self-interaction potential may induce a first order phase transition, producing gravitational waves from bubble collisions. We dubbed such a new scenario {\it violent Majoron model}, because it would be associated to a violent phase transition in the early Universe. Sphaleron constraints can be avoided if the global $U(1)_{B-L}$ is broken at scales lower than the electroweak scale, provided that the B-L spontaneously breaking scale is lower than $10\, {\rm TeV}$ in order to satisfy the cosmological mass density bound. The possibility of a sub-electroweak phase transition is practically unconstrained by cosmological bounds and it may be detected within the sensitivity of next generation of gravitational waves experiments: eLISA, DECIGO and BBO. We also comment on the possible detection in CEPC collider, where Majorons's production can be observed from Higgs' portals in missing transverse energy channels. 

\end{abstract}



\maketitle

\section{Introduction}
\label{s.intro}

\noindent 
The idea that the neutrino can be identified with its anti-particle, the antineutrino ($\nu=\nu^{c}$), was a wonderful intuition by Ettore Majorana in '37 \cite{Majorana37}. A Majorana mass term for the neutrino must imply a violation of the Lepton number of two units, i.e. $\Delta L=2$. Nowadays, the only realistic test of Majorana's hypothesis is the neutrinoless double beta decay ($0\nu2\beta$). The Majorana mass term may be originated by a spontaneous symmetry breaking of a global $U(1)_{L}$ or $U(1)_{B-L}$ extending the Standard Model. This leads to the %interesting 
possibility that a new pseudo Nambu-Goldstone boson, dubbed {\it Majoron}, can be coupled with neutrinos and be emitted in the $0\nu\beta\beta$-process \cite{Chikashige:1980ui,Gelmini:1980re,Schechter:1981cv}. Majorons have phenomenological implications not only in $0\nu\beta\beta$ experiments, but they can be limited by astrophysical stellar cooling processes and cosmological bounds. In particular, the spontaneous symmetry breaking scale of $U(1)_{L}$ or $U(1)_{B-L}$ is highly  constrained to be higher than the electroweak scale \cite{22a,22b,24a,24b,Cline:1993ht}. 
On the other hand, it was argued that such a VEV scale cannot be higher than $10\, {\rm TeV}$ for a cosmologically consistent Majoron model \cite{Akhmedov:1992hi}. 

%\begin{figure}[t]
%\centerline{ \includegraphics [height=6cm,width=1.1\columnwidth]{AAAMMAJORON5.pdf}}
%\vspace*{-1ex}
%\caption{We displayed cosmological bounds on Majoron couplings with neutrinos 
%and energy scale of $B-L$ symmetry breaking compared --- light and dark green regions ---
%with future limits from next gravitational waves experiments (eLISA, U-DECIGO and BBO) --- covered regions are under the black lines. The perturbativity bound on the Majoron coupling with neutrinos is also displayed --- in red line. The level curves $1,2,3,4,5$ are plotted at fixed scales of the combined parameter $C(v_{BL})$.  }
%\label{plot}  
%\end{figure}

Despite these considerations, the Majoron particle remains very elusive, with not so many direct detection channels being predicted. Nonetheless, the possibility of testing first order phase transitions (F.O.P.T) in the early Universe seems to be more promising after the recent discovery of gravitational waves (GW) at LIGO \cite{Caprini:2015zlo,Kudoh:2005as}. In particular, next generations of interferometers like eLISA and U-DECIGO will be also fundamentally important to test gravitational signal produced by Coleman bubbles from FOPT. The production of GW from bubble collisions was first suggested in Refs.~\cite{Witten:1984rs,Turner:1990rc,Hogan:1986qda,Kosowsky:1991ua,Kamionkowski:1993fg}.

New experimental prospectives in GW experiments have motivated a {\it revival} of these ideas in the context of new extensions of the Standard Model \cite{Schwaller:2015tja,Huang:2016odd,Artymowski:2016tme,Dev:2016feu,Katz:2016adq,Huang:2017laj,Baldes:2017rcu,Chao:2017vrq,Addazi:2016fbj,Ghorbani:2017jls,Tsumura:2017knk,
Huang:2017rzf}. In other words, the GW data may be used to test new models of particle physics beyond the standard model. 

With this paper we suggest to test the Majoron self-interactions and the details of the spontaneous symmetry breaking mechanism of the Lepton symmetry from GW interferometers. This is still a completely open possibility, related to a {\it first order phase transition} (F.O.P.T) of the Lepton symmetry in the early Universe. In particular, we show how next generation of interferometers like (e)LISA, U-DECIGO and BBO can test the spontaneous symmetry breaking scale $V_{BL}=10\, {\rm GeV}\div 10\, {\rm TeV}$. We dub such a Majoron particle associated to the F.O.P.T. a {\it violent Majoron}. We will discuss how the current limits %on the {\it violent Majoron} 
from LHC in missing transverse energy channels do not exclude the possibility of a violent Majoron. We emphasize that in the framework of the violent Majoron model, the CEPC collider will be able to test an energy scale overlapping the one associated to the eLISA sensitivity for GW signals from F.O.P.T.

\section{The model}
\noindent
We consider the extension of the Standard Model described by the gauge groups $SU_{c}(3)\times SU(2)_{L}\times U(1)_{Y}\times U(1)_{B-L}$, in which the lepton and baryon numbers are promoted to a $U(1)_{B-L}$ global symmetry\footnote{The implication of the Majoron in neutron-antineutron transitions were recently discussed in Refs.~\cite{Berezhiani:2015afa,Addazi:2015pia,Addazi:2015ata,Addazi:2016rgo}. }.
We introduce a complex scalar field coupled to neutrinos and to the Higgs boson that spontaneously breaks the $U(1)_{L}$ symmetry:
\begin{equation} \label{fff}
\mathcal{L}_{M}=fH\bar{L}\nu_{R}+h\sigma \bar{\nu}_{R}\nu_{R}^{c}+h.c.+V(\sigma,H)\,,
\end{equation}
where $h,f$ are Yukawa matrices of the model, and
\begin{equation}
\label{VsH}
V(\sigma,H)=V_{0}(\sigma,H)+V_{1}(\sigma)+V_{2}(h,\sigma)
\end{equation}
the potential, with 
\begin{eqnarray}
\label{VsH2}
V_{0}(\sigma,H)=&\ \ \lambda_{s}\left(|\sigma|^{2}-\frac{v_{BL}^{2}}{2}\right)^{2}+\lambda_{H}\left(|H|^{2}-\frac{v^{2}}{2} \right)^{2}\nonumber\\
&+\lambda_{sH}\left(|\sigma|^{2}-\frac{v_{BL}^{2}}{2}\right)\left(|H|^{2}-\frac{v^{2}}{2} \right)\,,
\end{eqnarray}
and higher order terms 
\begin{equation}
\label{Vos}
V_{1}(\sigma)=\frac{\lambda_{1}}{\Lambda}\sigma^{5}+\frac{\lambda_{2}}{\Lambda}\sigma^{*}\sigma^{4}+\frac{\lambda_{3}}{\Lambda}(\sigma^{*})^{2}\sigma^{3}+h.c.
\end{equation}
and
\begin{eqnarray}
\label{Vdhs}
V_{2}(H,\sigma)=&\ \ \beta_{1}\frac{(H^{\dagger}H)^{2}\sigma}{\Lambda}+\beta_{2}\frac{(H^{\dagger}H)\sigma^{2}\sigma^{*}}{\Lambda}\nonumber\\
&+\beta_{3}\frac{(H^{\dagger}H)\sigma^{3}}{\Lambda}+h.c.\, .
\end{eqnarray}

In principle, the scales of new physics entering non-perturbative operators may be different at each others. For convention, we  parametrize their differences in the couplings $\lambda_{i},\beta_{i}$.

When $\sigma$ gets a VEV, 
it may be decomposed in a real and a complex field:
\begin{equation}
\label{sigma}
\sigma=\frac{1}{\sqrt{2}}\left(v_{BL}+\rho+i\chi\right)\,.
\end{equation}
After the global $U(1)_{L}$ 
symmetry breaking, the RH neutrino acquires a Majorana mass term 
\begin{equation}
\label{Mhvd}
M=2hv_{BL}
\end{equation}
and a Dirac mass for the LH neutrino 
\begin{equation}
\label{mhvv}
m=\frac{1}{\sqrt{2}}fv\,,
\end{equation}
where $|\langle H\rangle|=v$ and $v/\sqrt{2}=174\, \rm GeV$.

The seesaw relations are obtained for $M>\!\!>m$, namely 
\begin{equation}
\label{Nnunu}
N=\nu_{R}+\nu_{R}^{c}+\frac{m}{M}\left(\nu_{L}+\nu_{L}^{c}\right)\,,
\end{equation}
\begin{equation}
\label{nunu}
\nu=\nu_{L}+\nu_{L}^{c}-\frac{m}{M}\left(\nu_{R}^{c}+\nu_{R}\right)\,,
\end{equation}
i.e.
\begin{equation}
\label{mNNN}
m_{N}\simeq M,\qquad m_{\nu}\simeq \frac{m^{2}}{M}\,,
\end{equation}
$m_N$ standing for the mass of the right-handed neutrino $N$. The Majoron corresponds to the pseudo-scalar field $\chi$, which is the Nambu-Goldstone boson of the spontaneously broken $U(1)$ symmetry, while the real scalar $\rho$ gets a mass $m_{\rho}\sim O(1)\,v_{s}$ once the self-coupling is assumed to be $O(1)$. 

In the Majoron model higher order terms, like the one entering $V_{1}$, are desired in order to induce a mass contribution that would not be allowed at perturbative level. For example, these may be induced either by gravitational effects \cite{Akhmedov:1992hi} or by exotic instantons (see e.g. Ref.~\cite{Addazi:2014ila}) at lower scales than the Planck scale\footnote{However, related theoretical aspects within the context of string theory are not completely understood. For instance a global $U(1)_{L}$ might be thought as a local $U(1)_{L}$ from {\it flavor branes}, since in string theory no exact global $U(1)$ symmetry can arise. This possibility seems highly bounded by the weak gravity conjecture, pointed out recently in Ref.~\cite{Addazi:2016ksv}.}.

\section{Gravitational Waves signal from Majorons}
\noindent
The spontaneous symmetry breaking of the $U(1)_{B-L}$ can be catalyzed by a first order phase transition (F.O.P.T). This induces the generations of Coleman's bubbles expanding at high velocity, which generate a stochastic cosmological background of gravitational radiation. Gravitational waves are generated by three main processes: i) bubble-bubble collisions; ii) turbulence induced by the bubble's expansion in the plasma; iii) sound waves induced by the Bubble's running in the plasma. The peak frequency of the GW signal produced by bubble collision has a frequency 
\begin{equation}
\label{nucollision}
f_{collision}\simeq 3.5 \times 10^{-4}\! \left(\frac{\beta}{H_{*}}\right)\!\! \left( \frac{\bar{T}}{10\, {\rm GeV}}\right)\! \left(\frac{g_{*}(\bar{T})}{10}\right)^{1/6}\!\!\!\! {\rm mHz},
\end{equation}
in which $\beta$ is related to the size of the bubble wall and is expressed in \eqref{deba}, $\bar{T}$ is the temperature at the F.O.P.T., $g_{*}(\bar{T})$ label degrees of freedom involved and the GW intensity is estimated as follows
$$\Omega_{collision}(\nu_{collision})\simeq $$
\begin{equation}
\label{CCC}
k\mathcal{E}^{2}\left(\frac{\bar{H}}{\beta}\right)^{2}\left(\frac{\alpha}{1+\alpha}\right)^{2}
\left( \frac{V_{B}^{3}}{0.24+V_{B}^{3}}\right)\left(\frac{10}{g_{*}(\bar{T})}\right).
\end{equation}
The coefficient $k$ introduced above has a numerical value 
$k\simeq 2.4\times 10^{-6}$, 
while 
\begin{equation}
\label{E}
\mathcal{E}(\bar{T})=\left[T\frac{dV_{eff}}{dT}-V_{eff}(T)\right]_{T=\bar{T}},
\end{equation}
\begin{equation} \label{a}
\alpha=\frac{\mathcal{E}(\bar{T})}{\rho_{rad}(\bar{T})},\ \ \ \ \rho_{rad}=\frac{\pi^{2}}{30}g_{*}(T)T^{4}.
\end{equation}

In \eqref{a} $\rho_{rad}$ stands for the radiation energy density, while $\bar{T}\simeq v_{BL}$ is the first order phase transition temperature, defined by
\begin{equation} \label{deba}
\beta=-\left[\frac{dS_{E}}{dt}\right]_{t=\bar{t}}\simeq \left[\frac{1}{\Gamma}\frac{d\Gamma}{dt}\right]_{t=\bar{t}},
\end{equation}
in which 
$$S_{E}(T)\simeq \frac{S_{3}(T)}{T},\qquad \Gamma=\Gamma_{0}(T){\rm exp}[-S_{E}(T)],$$
$$\Gamma_{0}(T)\sim T^{4},\quad S_{3}\equiv \int d^{3}r\left(\partial_{i}\sigma^{\dagger}\partial_{i}\sigma+V_{eff}(\sigma,T)\right).$$
$V_{B}$ represents the velocity of the bubble. The various values of $V_{B}$ will determine the amount of corrections from turbulence and sonic waves discussed later. 

The effective potential is the model dependent part of the Eq.(\ref{nucollision}). 
In particular the effective potential get thermal corrections 
which can be treated in the same approximation 
performed in Ref.~\cite{Delaunay:2007wb}:
\begin{equation}
\label{Veff}
V_{eff}(s,T)\simeq CT^{2}(\sigma^\dagger \sigma)+ V(\sigma,H)\,,
\end{equation}
where 
\begin{equation}
\label{coeff}
C=\frac{1}{4}\left(\frac{m_{\sigma}^{2}}{v_{LB}^{2}}+\lambda_{sH}+h^{2}-24K_{BL}\right)\,,
\end{equation}
with 
\begin{equation}
\label{coeff2}
K_{BL}=(\lambda_{2}+\lambda_{3})\frac{v_{BL}}{\Lambda}+\beta_{2}\frac{v_{BL}}{\Lambda}\,.
\end{equation}

The case of $K_{BL}\simeq 4\times 10^{-2}$ corresponds to the one testable by eLISA, U-DECIGO and BBO. From Eq.~\eqref{coeff2}, assuming $\lambda_{2,3}, \beta_{2}=1$, $v_{BL}\simeq 1\div 100\, {\rm GeV}$
and 
$$\frac{1}{4}\left[\frac{m_{\sigma}^{2}}{v_{LB}^{2}}+\frac{1}{4}\lambda_{sH}+h^{2}\right]\simeq 1$$ 
this corresponds to a scale 
of  
\begin{equation}
\Lambda\simeq 400\,{\rm GeV}\div 4\, {\rm TeV}\,, \nonumber
\end{equation}
while the GW signal is among $10^{-5}\div 10^{-3}\,{\rm Hz}$.
The effect of the Higgs as a dynamical particle are suppressed 
as $O(v_{BL}^{2}/v^{2})$, which are totally negligible for $v_{BL}\simeq 1\div 10\, {\rm Gev}$
compared to $O(1)$ uncertanties from from Bubble collisions and expansion details.

Let us remark that in principle other contributions from turbulence and sound waves may affect the estimate of the new physics scale by a factor $O(1)$, since at least they will affect the power spectrum density of GW by a factor $O(10)$. This may lower the scale of the new physics by a factor $3$. 

Let us compere these order of magnitude semi-analytical estimations 
with numerical simulations. 
In Fig.1, we show numerical plots in a realistic set of parameters,
using the same model independent spectrum parameterization 
of Ref.\cite{Caprini:2015zlo}. We also consider the contribution of 
turbulence and shock waves as in Ref.\cite{Caprini:2015zlo}.
The results are in good $O(1)$ agreement with the estimations inferred above. 

\begin{figure}[t]
\centerline{ \includegraphics [height=4.3cm,width=0.9\columnwidth]{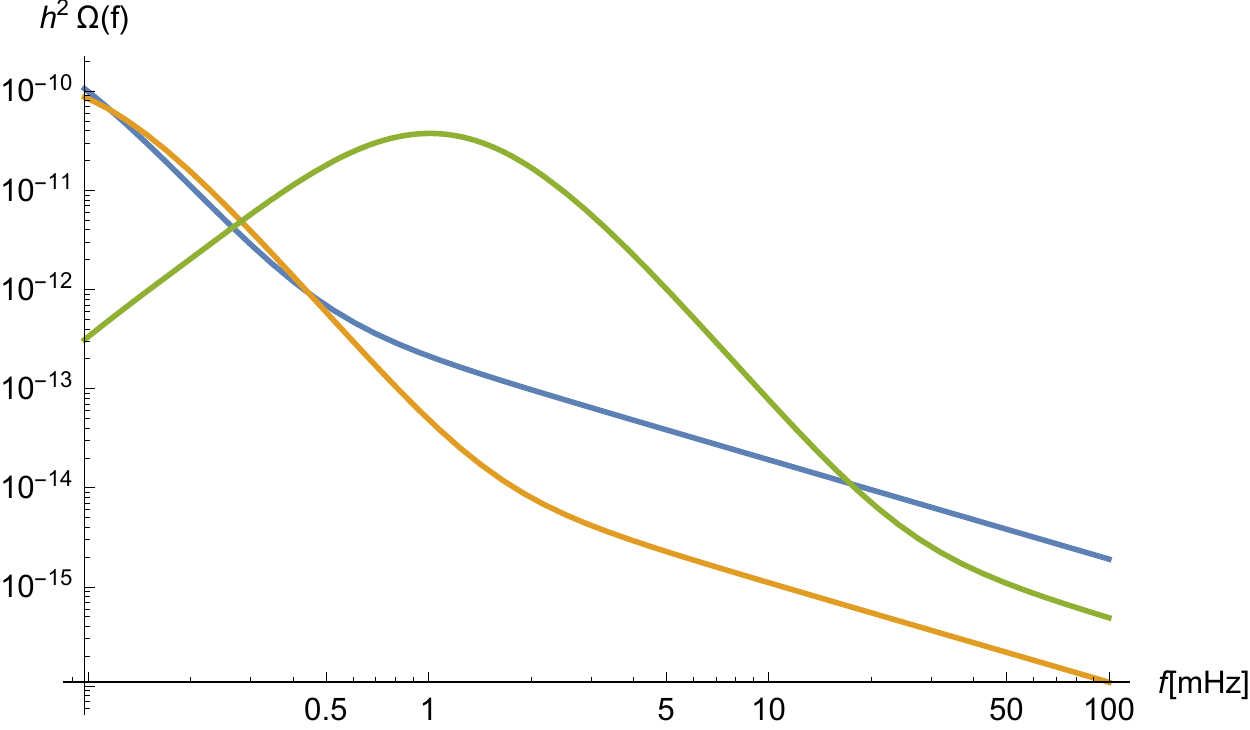}}
\vspace*{-1ex}
\caption{The gravitational waves energy density as a function of the frequency is displayed.
We use the same model independent parametrization of Ref.\cite{Caprini:2015zlo}. 
We show three {\it non-runnaway} bubbles cases which are compatible with the B-L first order phase transition: 
In blue, we consider the case of 
$\bar{T}=50\, {\rm GeV}$, $\beta/\bar{H}=100$, $\alpha=0.5$, $\alpha_{\infty}=0.1$, $V_{B}=0.95$;
in green $\bar{T}=20\, {\rm GeV}$, $\beta/\bar{H}=10$, $\alpha=0.5$, $\alpha_{\infty}=0.1$, $V_{B}=0.95$.
Orange: $\bar{T}=10\, {\rm GeV}$, $\beta/\bar{H}=10$, $\alpha=0.5$, $\alpha_{\infty}=0.1$, $V_{B}=0.3$. 
The three cases lies in the sensitivity range of LISA \cite{Caprini:2015zlo}. }
\label{plot}   
\end{figure}

\subsection{LHC constrains}
\noindent
From LHC important constrains on the Higgs decay into invisible channels are set. Let us define
\begin{equation}
\label{mu}\mu_{F}=\frac{\sigma^{NP}(pp\rightarrow H)}{\sigma^{SM}(pp\rightarrow H)}\frac{BR^{NP}(H\rightarrow F)}{BR^{SM}(H\rightarrow F)}\,,
\end{equation}
where $F=\gamma\gamma, WW,ZZ,\tau,\tau$ label final states. In Tab/Fig.~1 we show the limits from various channels on Higgs decays. Comparing Eq.(\ref{mu}) with limits from LHC in Fig.~2, we can set a bound on the $C_{H\chi\chi}$-parameter in the Higgs decay rate. 

\begin{figure}[t]
\centerline{ \includegraphics [height=4.3cm,width=0.9\columnwidth]{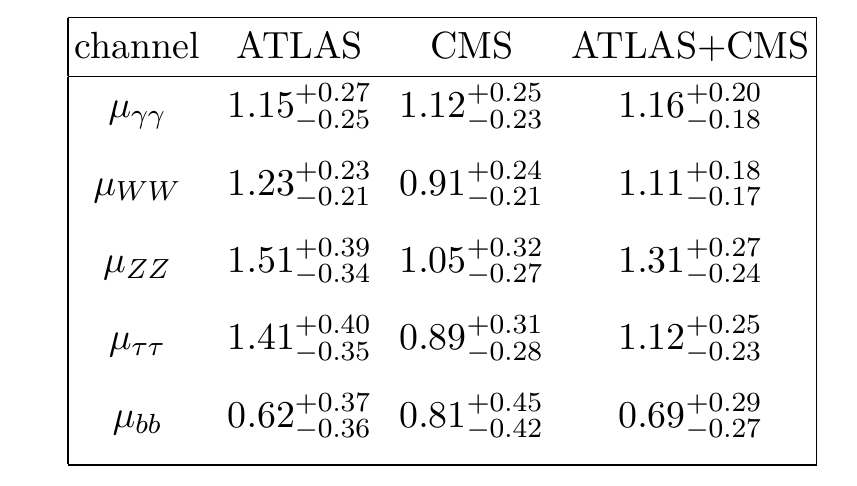}}
\vspace*{-1ex}
\caption{Combined bounds on the $\mu$-parameters associated to the Higgs decays into 
SM channels.}
\label{plot}   
\end{figure}

In particular, the model independent limit placed on the invisible decays branching ratio 
is (see e.g. \cite{Chatrchyan:2014tja})
\begin{equation}
\label{Hinvisible}
{\rm Br}(H\rightarrow invisible)=\frac{\Gamma_{inv}}{\Gamma_{inv}+\Gamma_{SM}}<0.51\,(95\%\, {\rm C.L.}), 
\end{equation}
which corresponds to the bound
\begin{equation}
C_{H\chi\chi}\leq 0.6\,.
\end{equation}
This amounts to find a maximal bound on the hierarchy between the $v_{\sigma}$ and $\Lambda$ scale that corresponds to the value 
\begin{equation}
C_{H\chi\chi}\simeq C_{H\chi\chi}^{0}=\frac{\beta_{2}v_{BL}}{\Lambda}<0.6\,,
\end{equation}
where $C_{H\chi\chi}^{0}$ is the leading order contribution to the $C_{H\chi\chi}$, 
which is originated from the operator in Eq.(\ref{Vdhs}) parametrized by $\beta_{2}$. 
Such a constrain is easily compatible with GW signals in eLISA and cosmological bounds. 
For example, fixing $v_{BL}=100\, {\rm GeV}$ and $\beta_{2}=1$, $\Lambda>166\, {\rm GeV}$ is enough to avoid LHC constraints,
while for eLISA $\Lambda\simeq 4\, {\rm TeV}$ is large enough to generate a detectable GW signal
-- see the previous section on gravitational waves discussed above.

%\begin{figure}[t]
%\centerline{ \includegraphics [height=6cm,width=0.7\columnwidth]{NVIOLENT1A.pdf}}
%\vspace*{-1ex}
%\caption{We report the limits from LHC and future CEPC 
%(in  brown and blu respectively) and the cosmological sphaleron bounds (green). 
%}
%\label{plot}   
%\end{figure}
\begin{figure}[t]
\centerline{ \includegraphics [height=6cm,width=0.7\columnwidth]{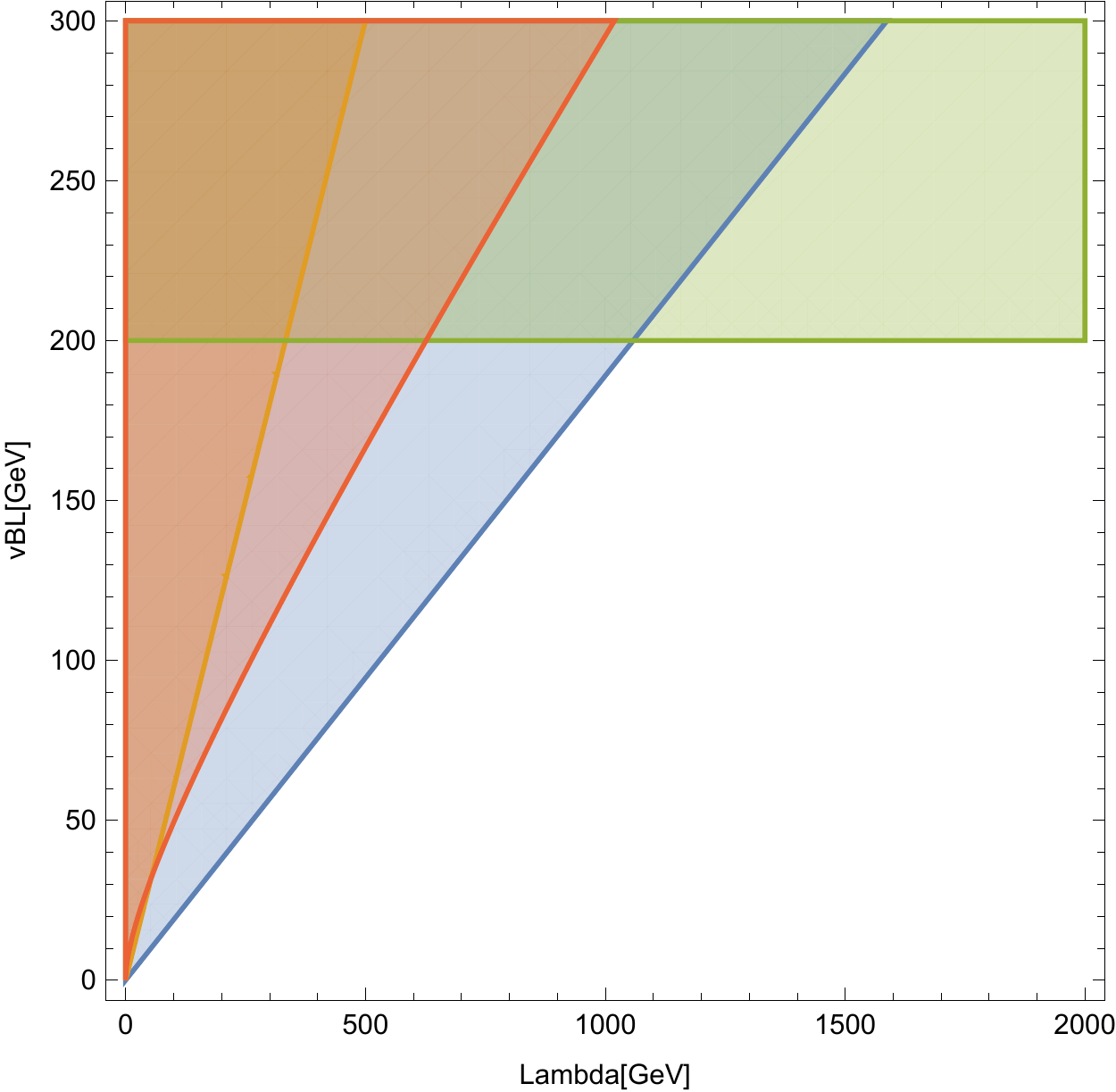}}
\vspace*{-1ex}
\caption{We report the limits from LHC and future CEPC 
(in  brown and blu respectively), cosmological sphaleron bounds (green) and the region which will be probed by eLISA (red). 
The case of $\beta_{2}=1$ is displayed. }
\label{plot}   
\end{figure}

\subsection{electron-positron colliders and invisible Higgs decays}
\noindent 
In CEPC, the Higgs decay rate will be probed with a factor $O(5\div 10)$ of sensitivity higher than LHC. 

The golden channel will be the process 
\begin{eqnarray}
\label{Process}
e^{+}e^{-}\rightarrow ZH\rightarrow Z\bar{b}b\,, \nonumber
\end{eqnarray}
with a cross section
\begin{eqnarray}
\sigma_{hZ\rightarrow b\bar{b}Z}=\sigma_{hZ}^{SM}\times R_{hZ}\times BR(h\rightarrow b\bar{b})\,, \nonumber
\end{eqnarray}
where $R$ is a suppression factor related to the coupling of the Higgs boson to $Z$. 
According to Ref. \cite{TLEP}, 
the number of Higgs boson produced at Center of Mass energy $\sqrt{s}=240\, {\rm GeV}$ from the $e^{+}e^{-}\rightarrow HZ$ channel
will be $2\times 10^{6}$, higher then the other channels. 
The total cross section of the process can be measured with $0.4\%$ of precision,
while $\sigma_{HZ}\times Br(h\rightarrow b\bar{b})$ can reach precision $0.2\%$ for $\sqrt{s}=240\, {\rm GeV}$ \cite{TLEP},
while it can reach one order of magnitude more then LHC in the $\sqrt{s}=350\, \rm GeV$ (see Fig. 12 of  \cite{TLEP}).
In Fig.~3 we consider the constraints on Eq.~(\ref{Hinvisible}) overimposed to current LHC constrains, cosmological bounds and future eLISA sensitivity regions. 

Let us conclude this section with an important remark. 
We would like to stress again that the  Fig.2 was displayed 
fixing  $\beta_{2}$  to be one. 
However, fixing different values of the
$\beta_2$ parameter 
will change the region plot shown.
 Assuming smaller values of $\beta_{2}$,  the constrain on the $\Lambda$ from colliders are relaxed. Let us also note that the case $\beta_{2}<<1$ is still possible. This case will correspond to suppress the invisible channels in colliders, rendering the Gravitational interferometers bounds stronger then colliders one:
the gravitational power spectrum depends on a large set of initial parameters $\lambda_{1,2,3},\beta_{1,2,3}$.
With $\beta_{2}<<\lambda_{1,2,3},\beta_{1}$, the GW signal is still unsuppressed. 
It is worth to mention that 
 for $\beta_{2}\simeq 0$, Invisible channels can be generated from radiative corrections, but, of course, with a strong suppression factor.

\subsection{Cosmological limits}

\subsubsection{Sphaleron bounds}

Relating the Majoron model to the pre-sphaleron leptogenesis, 
stringent constrains are provided from washing-out processes of $(B+L)$-violating sphaleronic interactions in the Standard Model. 

First of all, we remind that the bound on the cosmological neutrino mass is 
\begin{equation}
\label{neutrino}
m_{\nu}\lesssim 50\, {\rm KeV}\left(\frac{100\, {\rm GeV}}{T_{BL}}\right)^{1/2}\,,
\end{equation}
where $T_{BL}$ is the temperature at which the $L$ (or $B-L$) asymmetry is generated.
The cosmological neutrino mass bound sets in turn a bound on $T_{BL}$ from Eq.~(\ref{neutrino}) that reads  
\begin{equation}
\label{mnu}
m_{\nu}\lesssim 10^{-3}\, {\rm eV}\rightarrow T_{BL}\simeq 10^{12}\, {\rm GeV}\,. 
\end{equation}
The neutrino mass bound used in all estimations of these papers consider the cosmological 
and neutrinoless double beta decay bounds (see Ref. \cite{PDG}). 
From sphalerons constraints, one can get out the following bound on $v_{s}$, i.e.
\begin{equation}\label{vs}
v_{s}\lesssim {\rm Max}\left( 200\, {\rm GeV}\lambda_{1}^{-1/7}(27\,Y)^{-4/7}U_{hl}^{-8/7},v\right)\,,
\end{equation}
where
\begin{equation}
\label{Y}
Y=\frac{n_{\chi}}{n_{\gamma}}\,,
\end{equation}
which reduces to $Y\simeq 1/27$ \cite{22a,22b}, is compatible with Big Bang Nucleosynthesis constraints \cite{24a,24b}, and $U_{hl}$ represents the neutrino mixing matrix. Eq.~(\ref{vs}) can be related to a bound on the Yukawa matrices:
\begin{eqnarray}
\label{minl}
\!\!\!\!\!\!\!\!\!{\rm min}_{l}\sum_{i}\frac{|h_{il}|^{2}}{f_{2i}}<&&6\times 10^{-14} \times \nonumber\\
&&{\rm Max}\left(1,0.8\lambda^{-1/7}U_{hl}^{-8/7}(2Y)^{-4/7} \right)\,.
\end{eqnarray}
Eq.~(\ref{minl}) provides a very strong bound on $h$-couplings that reads
\begin{equation}
\label{hil}
h_{il}\lesssim 10^{-7}.
\end{equation}
This bound is so strong to be valid even for tiny gravitationally induced (Planck scale suppressed) effects, or when the mixing $U_{hl}$ are very small.  
Finally, let us comment on possible way-out to this bound.
The sphaleron bound can be relaxed, allowing $v_{BL}>v_{Higgs}$, 
in electroweak baryogenesis scenarios
and post-sphaleron baryogengesis scenarios. 
In this case, also gravitational waves signal  from the baryogengesis scenarios should be observable
-- as recently discussed in Ref. \cite{Huang:2016odd}.

\subsubsection{Cosmological density bound}

Cosmological density constraints can be set distinguishing the two cases:
\begin{equation}
\label{AB}
(A)\,\,\,v_{BL}<v,\qquad (B)\,\,\,v_{BL}>v.
\end{equation}
In the case 
(A), the Majoron mass is dominated by the $\beta_{1}$-term
and casts 
\begin{equation}
\label{mass}
m_{\chi}\simeq \beta_{1}^{1/2}\left(\frac{v}{v_{BL}}\right)^{1/2}\,{\rm KeV}. 
\end{equation}

In the case (B) the mass of the Majoron reads
\begin{equation}
\label{masschi}
m_{\chi}\simeq \left(\frac{25}{2}\lambda_{1}+\frac{9}{2}\lambda_{2}+\frac{1}{2}\lambda_{3}\right)^{1/2}\left(\frac{v_{BL}}{v}\right)^{3/2}\, {\rm KeV}\,,
\end{equation}
with
\begin{equation}
\label{h}
h\simeq \sqrt{2}\frac{m_{\nu}^{D}}{v}\simeq \frac{\sqrt{2m_{\nu}M}}{v}\,,
\end{equation}
where $M$ is the RH neutrino mass while $v$ is the Higgs VEV. The cosmological density 
constraint on the Majorons reads
\begin{equation}
\label{nm}
n_{\chi}m_{\chi}<\rho_{crit}\,,
\end{equation}
for $n_{\chi}\simeq n_{\gamma}$ and for $v=v_{BL}$ and $\lambda,\beta<10^{-2}$. On the other hand, for $\chi$ decoupling sufficiently rapidly, it should be possible that $n_{\chi}<\!\!<n_{\gamma}$. Consequently the constraints on $\chi$ can be weaker than the above. 

Cosmological constraints that can be derived from Majorons heavily rely on the assumption  that the Majoron is out of equilibrium, and on the decay channels that are allowed by the particular model instantiations. 

If we consider massive Majorons and stable LH neutrinos, limits on the Yukawa coupling $h$ can be derived from the see-saw relation and the cosmological constraints on the neutrinos mass density, i.e.
\begin{equation}
\label{mnu}
h\simeq \frac{\sqrt{2m_{\nu}M}}{v}\leq 10^{-6}\left(\frac{M}{{\rm GeV}}\right)^{1/2}\,,
\end{equation}
where $v$ is the Higgs expectation value. LH electrons and RH neutrinos are in thermal equilibrium via the interactions
\begin{equation}
\label{psinuW}
\psi_{L}+h\rightarrow \nu_{R}+W_{L}\,,
\end{equation}
with an interaction rate 
\begin{equation}
\label{int}
\Gamma\simeq \frac{g^{2}h^{2}}{16\pi}T.
\end{equation}
Thermal equilibrium, as realized above, happens for 
\begin{equation}
\label{neutrinos}
M \leq T\leq 10^{5}M\,.
\end{equation}
For $T<M$, RH neutrinos go out of thermal equilibrium, disappearing from the thermal bath. At this stage, the relevant interaction of the scalar complex field $\sigma$ is with LH neutrinos, with a coupling of the order of 
\begin{equation}
\label{BL}
f\simeq \frac{m_{\nu}}{v_{BL}}\,.
\end{equation}
In the case of a spontaneous symmetry breaking scale of $v_{BL}\simeq 1\div 100\, {\rm GeV}$ suggested above while discussing GW signals, the limit on the Majoron coupling becomes very stringent and reads  
\begin{equation}
\label{f}
f\simeq 2\times (10^{-8}\div 10^{-10})\,.
\end{equation}

Let us remark that, assuming $M>10\, {\rm GeV}$ os so, $\sigma$ goes out of equilibrium for a temperature of about $T\simeq M$.  As a consequence, Majoron density in the present Universe is 
\begin{equation}
\label{rchi}
r_{\chi}=\frac{n_{\chi}(T_{0})}{n_{\gamma}(T_{0})}=\frac{g_{*}(T_{0})}{g_{*}(T_{RH})}\simeq 0.1\div 0.2\,,
\end{equation}
in which $g_{*}(T)$ represents the effective number of light particle species at a temperature $T$; $T_{RH}$ and $T_{0}$ are the decoupling temperature for the RH neutrinos and the presente temperature of the Universe respectively. 

Further constrains on the $v_{BL}$ from Majoron decays must be considered. The Majoron decay into two neutrinos 
\begin{equation}
\label{chi}
\chi\rightarrow \nu\nu
\end{equation}
has a decay time 
\begin{equation}
\label{tau}
\tau_{\chi}=8\pi \left(\frac{v_{BL}}{m_{\nu}}\right)^{2}m_{\chi}^{-1}.
\end{equation}
Eq.~(\ref{tau}) constrains the Majoron relic density as follows  (see Ref. \cite{Akhmedov:1992hi}): 
\begin{equation}
\label{rchi}
r_{\chi}m_{\chi}\left(\frac{\tau_{\chi}}{\tau_{U}}\right)^{1/2}<25(\Omega_{0}h^{2})\,{\rm eV}\,,
\end{equation}
where $\tau_{U}$ is the age of the Universe. Eq.~(\ref{rchi}) leads to 
\begin{equation}
\label{rchichi}
r_{\chi}\left(\frac{m_{\chi}}{{\rm keV}} \right)\left(\frac{\tau_{\chi}}{{\rm sec}} \right)^{1/2}
\leq 10^{7}\Omega_{0}h^{3/2}\,,
\end{equation}
where $\delta \rho/\rho\leq 10^{-4}$ are the initial density fluctuations. \\

On the other hand relativistic decay products of $\chi$ must be redshifted enough 
to maintain a matter dominated Universe, i.e. to avoid constrains on dark radiation: 
\begin{equation}
\label{tchichi}
t_{\chi}n_{\gamma}(t_{eq})m_{\chi}\left(\frac{\tau_{\chi}}{t_{eq}}\right)^{1/2}<\rho_{m}(t_{eq})\,,
\end{equation}
where 
\begin{equation}
\label{tchichich}
n_{\gamma}(t_{eq})=(1+z_{eq})^{3}\times 422\,{\rm cm}^{-3}
\end{equation}
and
\begin{equation}
\label{tchichichd}
\rho_{M}(t_{eq})=(1+z_{eq})^{3}\times 10.5(\Omega_{0}h^{2})\, {\rm KeV\, cm^{-3}}\,.
\end{equation}
This leads to the following bound
\begin{equation}
\label{mchi}
m_{\chi}\left( \frac{v_{BL}}{v}\right)^{2}\leq 10^{6}\left(\frac{m_{\nu}}{25\, {\rm eV}}\right)^{2}\, {\rm keV},
\end{equation}
leading to 
\begin{equation}
\label{vBLBL}
v_{BL}<\left(\frac{m_{\nu}}{25\,{\rm eV}}\right)^{4/7}\times 10\, {\rm TeV}. 
\end{equation}
Such a bound can be generalized 
for higher order operators in the complex scalar sector, leading to
\begin{equation}
\label{sigma}
\sigma^{4+n}/\Lambda^{n} \rightarrow v_{BL}<10^{10/(n+6)}\left(\frac{\Lambda}{{\rm GeV}}\right)^{n/(n+6)}\! {\rm GeV}\,.
\end{equation}
Nonetheless, bounds on $v_{BL}$ provided by higher $n$-order contributions are less stringent. 

\subsubsection{Dark Matter}
The Majoron particle can provide a new candidate of Dark Matter \cite{Akhmedov:1992hi,DM}. 
For sub-electroweak phase transition considered, 
the Majoron mass is parametrized by Eq.(\ref{mass}). 
For a $v_{BL}\simeq 10\, {\rm GeV}$ phase transition, 
the Majoron is naturally Kev-ish, 
while the overproduction problem is avoided. 
 This means that the Majoron can compose Warm Dark Matter. 
The Majoron Dark Matter paradigm can be tested from colliders missing energy channels 
and from gravitational waves experiments. This certainly enforces 
the naturalness and phenomenological health of our proposal. 

\noindent

\medskip

\noindent

\section{Conclusions and remarks}
\label{s.disc}
\noindent
We have shown how gravitational waves experiments may provide useful informations on the Majoron self-interaction potential. In particular, the possibility of a first order phase transition at a scale of about $1\div 100\, {\rm GeV}$ is still unbounded by any cosmological limits, such as non-perturbative electroweak effects --- sphalerons --- and cosmological density abundance. The main message of this paper is that such a scale overlaps the sensitivity of future gravitational waves' interferometers, like eLISA, U-DECIGO and BBO. In fact, a scale of about $1\div 100\, {\rm GeV}$ falls into the range of frequencies around $10^{-5}\div 10^{-3}\, {\rm Hz}$. However, an observation of a stochastic gravitational waves signal in eLISA should imply a new physics scale of UV completion for the violent Majoron model of about $3\, {\rm TeV}$ or so. This means that the production of Majorons in colliders may provide a complementary {\it test-bed} for this model. For instance, the detection of Majorons in missing energy channels can be tested in the future collider CPEC from Higgs invisible decays.

\noindent

\acknowledgments
\noindent 
AM~wishes to acknowledge support by the Shanghai Municipality, through the grant No. KBH1512299, and by Fudan University, through the grant No. JJH1512105.

\end{document}